\documentclass[11pt,draft]{article}
\usepackage{epsfig,amssymb,cite,multirow} 
\usepackage{amsmath}

\usepackage{amscd}
\usepackage{amsmath,graphicx}
\usepackage{xypic}
\usepackage[matrix,arrow,curve]{xy}

\usepackage{amsmath,amsthm}

\def \d {\mbox{d}}
\def \w {\wedge}

\def\del          {\partial}
\def\delbar       {\bar\partial}

\def\Re           {{\rm Re\hskip0.1em}}
\def\Im           {{\rm Im\hskip0.1em}}

\def\d         {{\rm d}}

\def\sqr#1#2{{\vcenter{\vbox{\hrule height.#2pt
 \hbox{\vrule width.#2pt height#1pt \kern#1pt \vrule width.#2pt}\hrule
 height.#2pt}}}}

\newcommand{\bbb}{{\cal B}}
\newcommand{\N}{{\cal N}}

\makeatletter
\@addtoreset{equation}{section}
\makeatother

\def\beq{\begin{equation}}                     %  
\def\eeq{\end{equation}}                       % 
\def\bea{\begin{eqnarray}}                     %         % 
\def\eea{\end{eqnarray}}   
\def\nn{\nonumber}                    %       %  
\begin{document}
\setcounter{page}{0}
\begin{titlepage}
\titlepage
%\rightline{hep-th/}
\rightline{SPhT-T09/109}
%\rightline{LPTHE-06-XX}
%\rightline{Bicocca-FT-09-YY}

\vskip 3cm
\centerline{{ \bf \Large New families of interpolating type IIB backgrounds}}
%\vskip 0.5cm
%\centerline{{ \bf \Large }}
\vskip 1.5cm
\centerline{Ruben
Minasian $^{a}$, Michela Petrini $^{b}$
and Alberto Zaffaroni $^c$}
\begin{center}
$^a$ Institut de Physique Th\'eorique,                   
CEA/Saclay \\
91191 Gif-sur-Yvette Cedex, France  
\vskip .4cm
$^b$ LPTHE, Universit\'e Paris VI, \\
4 place Jussieu, 75252 Paris, France
\vskip .4cm
$^c$ Universit\`a di Milano Bicocca and INFN, sezione di
Milano-Bicocca \\
piazza della Scienza 3, Milano 20126, Italy
\end{center}
\vskip 1.5cm  
\begin{abstract}

We construct new families of interpolating two-parameter solutions of type IIB  supergravity.
These correspond to D3-D5 systems on non-compact six-dimensional manifolds which are 
$\mathbb{T}^2$ fibrations over Eguchi-Hanson and multi-center Taub-NUT spaces, respectively.  
One end of the interpolation corresponds to a solution with only D5 branes and vanishing 
NS three-form flux. A topology changing transition  occurs at the other end,
where the internal space becomes a direct product of the four-dimensional surface and 
the two-torus and the complexified NS-RR three-form flux becomes imaginary self-dual.  
Depending on the choice of the connections on the torus fibre, 
the interpolating family has either $\N=2$ or $\N=1$ supersymmetry. In the $\N=2$ case it can be shown
that the solutions are regular.

\end{abstract}

\vfill
\begin{flushleft}
{\today}\\
%\vspace{.5cm}
\end{flushleft}
\end{titlepage}

\newpage
\large

%%%%%%%%%%%%%%%%%%%%%%%%%%%%%%%%%%%%%%%%%%%%%%%% 

%\centerline{\bf \Large Branes and fluxes on balanced noncompact six-manifolds}

\section{Introduction}

In recent years a lot of attention has been devoted to the study of supergravity backgrounds
with non trivial fluxes. Backgrounds of this kind are indeed relevant for both string
compactifications to four dimensions and the study of the gauge/gravity duality. In all cases,
the geometries of interest are of warped product type: a product of four dimensional
Minkowski (or Anti-de Sitter) space and an internal six-dimensional manifold, either 
compact or not.

A particular class of flux backgrounds is provided by solutions where the internal 
geometry is given by  a so-called $SU(3)$ structure manifold.
$SU(3)$ structure manifolds are the simplest generalisation of Calabi-Yau's, in that
they also possess a globally defined holomorphic three-form and a globally defined
fundamental form, but these, contrary to Calabi-Yau's, are not closed.
The non closure of such forms is a measure of the back-reaction of the non trivial
RR and/or NS fluxes. 

In the formalism of  generalized complex geometry it easy to observe some general features of 
type IIB backgrounds with $SU(3)$ structure: the internal manifold must be complex
and there is an intricate cancellation among $\d J$,  the exterior derivative of the fundamental 
form compatible with the integrable complex structure, and the NS and RR there fluxes,
$H$ and $F_3$ \cite{gmpt1,gmpt2}. 

One very special case is when the symplectic form is closed.
Then the internal manifold is conformally Calabi-Yau, and the three-form fluxes form
%which form  
an imaginary self-dual complex combination.
%, cancel each others contribution. 
%The warp factor determines the RR five-form. 
This situation is typically referred to as type B solution \cite{Mgp}. 
Another special point, which is referred as type C, is reached when $H=0$. 
The only non trivial flux  is the RR three-form, which is proportional to the exterior derivative of the fundamental form. The dilaton is running and proportional to the warp factor ($\phi = 2A$).
In a general solution, which is referred as of interpolating type,  the symplectic structure is not integrable. Taking in addition  a constant axion, one can show that the NS three-form flux is exact and hence trivial in cohomology. \\

%As far as the topological properties of the NS sector go, 
%the general solution and the type C point have very similar properties. In particular they 
%require a complex internal six-manifold which does not admit a K\"ahler metric.\\

The best known examples of gravity duals to $\mathcal{N}=1$ supersymmetric gauge theories,
the Klebanov-Strassler (KS) \cite{KS} (type B) and the Maldacena-Nunez (MN) \cite{MN} (type C) solutions, 
fall in the $SU(3)$ structure class of backgrounds. So does the interpolating family found in  \cite{bgmpz},
which describes both the gravity dual to the baryonic branch of the Klebanov-Strassler solution and an interpolation between  KS and MN. \\

Recently a class of torsional non-K\"ahler geometries satisfying the conditions  of  $\mathcal{N}=1$ supersymmetry
in purely NS backgrounds has been constructed in \cite{GP}, as a subclass of principal holomorphic $\mathbb{T}^2$ fibrations over a four-dimensional 
hyper-K\"ahler surface $\bbb$\footnote{Incidentally, the type C solution is S-dual to the third special point, type A, where $F_3 = 0$, 
and which is hence pure NS type. Since the supersymmetry conditions are local, forgetting about the tadpole, 
the same set of requirement on the internal space would apply to  purely NS background both in type II theories 
and in heterotic strings. In fact the local construction of the internal six-manifold that we use here \cite{GP}, 
as well as one of the explicit examples \cite{fty}  appeared originally in the heterotic context.}.
A generic feature of such backgrounds is that they admit a balanced metric, i.e. have a fundamental form whose square is closed. 
In this paper, we shall use the same kinds of torsional geometry to  construct type  IIB backgrounds.
Since we are interested in non-compact backgrounds, we shall work with non-compact base manifolds, $\bbb$,
and consider multi-center Taub-NUT surfaces; the two-center case, the Eguchi-Hanson space,  
will be treated in a greater detail. The construction of \cite{GP} is local, and one has to ensure that the Bianchi identities are satisfied. 
%We 
%will impose the requirement that the NS three-form flux is closed and the total RR flux is $H$-twisted closed. 
For the  holomorphic $\mathbb{T}^2$ fibrations over Eguchi-Hanson space,  
the solution of the Bianchi identity has been given in \cite{fty} for the heterotic string. For this special case,
we shall heavily rely on the analysis there.\\

In type IIB we find a  web of $\mathcal{N}=2$  backgrounds connected by different 
types of dualities and we discuss possible generalisations to $\mathcal{N}=1$.
The simplest solution is of type C  
%which is the S-dual of the purely NS  solution
and describes D5-branes wrapping a two-cycle in the internal six-manifold. 
Due to the smooth $\mathbb{T}^2$ action on the internal, the background has isometries and hence can be T-dualized. The result is a type B solution, with constant dilaton (and axion), where the internal space is a direct product $\bbb \times \mathbb{T}^2$ but the $B$ field is not vanishing, and $H$ is non-trivial in cohomology.  There is also a non-zero three-form, which is dual to the NS flux. 

These two solutions are  the respective counterparts of the MN and
KS backgrounds.  It is then natural to look for an interpolation connecting the two, in an analogy  to the gravity dual to the baryonic branch \cite{bgmpz}.
Recently, a method has been proposed, \cite{mm}, that allows, given any
type C solution, to build a new one of interpolating type.  The construction of \cite{mm} was motivated by a chain of dualities including an M-theory boost and was applied to reinterpret the interpolating solution of \cite{bgmpz}. An interesting feature of the construction is that all the dualities are ``external"; they do not change the structure of the internal manifold and simply modify 
the relation between the dilaton and the warp factor,
$$
e^{-(2A-\phi)} \sim \sin w \, ,
$$
to include the interpolating function $w$.
In view of the fact that there is no topology change between the type C and the interpolation, this 
construction is very natural. As we shall see all Bianchi identities in the general interpolating solution reduce to the type C Bianchi identity.
The interpolating solutions are in fact a two parameter family of solutions, where the parameters are
the M-theory boost and the asymptotic value of the dilaton. Notice that the existence of (at least) two parameters 
is another generic feature of type IIB interpolating families (with fixed axion). 

Since our background has isometries, we can obtain the interpolating solution also using 
``internal" dualities. We shall show that deforming a type B solution on  $\bbb \times \mathbb{T}^2$  by adding a closed $B$ field and then performing T-duality, we can again recover the same interpolation. 
%The standard imaginary seld-dual combination $F_3 + i e^{-\phi} H$ of type B solution is mapped into the imaginary self-duality of $F_3 + i \frac{1}{\cos w} \,e^{-\phi}H$. \\

Finally, since the solutions are non compact, it would be interesting to see whether they admit a holographic gauge dual and
how this looks like. A natural candidate would be a four-dimensional $\mathcal{N}=2$ gauge theory obtained by compactifying
on the two torus the little string theory dual to D5-branes transverse to the ALE space.  However we will not explore this in details in this paper.\\

The plan of the paper is as follows. 
Before moving to the discussion of the various solutions, we conclude the introduction
with a brief review of six-dimensional internal geometry and the supersymmetry
equations cast in the pure spinor form \cite{gmpt2}.
 In Sections 2 and 3 we describe the type 
C and its T-dual type B solutions, respectively. Then in Section 4 we present the interpolating solution  and discuss its properties.
We briefly comment about a possible holographic interpretation in the Conclusion.

\subsection{Torsional geometry}
\label{tg}

We are interested in supersymmetric solutions of type IIB supergravity of warped  type
\beq
\label{metric}
\mbox{d}s^2 = e^{2 A} \mbox{d}s_4^2 + \mbox{d}s_6^2  \, ,
\eeq
where the internal manifold is a six-dimensional complex non-K\"ahler manifold with $SU(3)$ structure. 
As discussed in \cite{GP}, manifolds of this kind can be obtained via $\mathbb{T}^2$ fibrations over a four-dimensional hyper-K\"ahler manifold, $\mathbb{T}^2 {\rightarrow} X \stackrel{\pi}{\rightarrow}\bbb$. 

Provided that  the $\mathbb{T}^2$ bundle is non-trivial and that its curvature has no components in $\Lambda^{0,2} T^*\bbb$, the space of fibrations
 will admit a closed holomorphic three-form, and hence an integrable complex structure. It will not admit a K\"ahler metric. Typically, a necessary condition for preserving supersymmetry is that the internal six-manifold admits a balanced metric, i.e. has a fundamental form whose square is closed. It  is satisfied by requiring that in addition, the curvature of the $\mathbb{T}^2$ bundle is orthogonal to the K\"ahler form on $\bbb$. Finally, if the curvature has no components in $\Lambda^{2,0} T^*\bbb$, i.e. lies completely the space of anti-self-dual two-form on the base  $\Lambda^{1,1-} T^*\bbb$, 
the supersymmetry will be enhanced to $\N=2$. \\

For the metric in the base space, $\mathcal{B}$, we take
\beq
\label{basemetric}
 \mbox{d}s_\bbb^2  = V \, (\d y_1^2 + \d y_2^2 + \d y_3^3) + V^{-1} (\d \tau + A)^2 \, ,
\eeq
where $V$ is a harmonic function and $A$ is a connection one-form $A= A_i \d y^i$. 
In order for the space to be  hyper-K\"ahler, $V$ has to satisfy the monopole condition
\beq
\label{monopole}
\partial_i V = -  \epsilon_{ijk} \partial_j A_k \qquad i=1,2,3 \, .
%\qquad (\d V = *_3 \d A) \, .
\eeq 

Indeed, choosing vielbein  $e^0 = V^{-\frac{1}{2}}  (\d \tau + A) $ and $e^i = V^{\frac{1}{2}} \d y^i$  ($i = 1,2,3$), it is not hard to see that the two-forms
\beq
\label{2forms}
\Omega^i_{\pm} = e^0 \wedge e^i \pm \frac{1}{2} \, \epsilon^i_{jk} e^j \wedge e^k
\eeq
are respectively self- and anti-self-dual ($* \Omega^i_{\pm} = \pm \Omega^i_{\pm} $). Moreover 
the  self-dual forms are closed:
\beq
\label{2forms-closed}
\d \, \Omega^i_{+} = 0 \, .
\eeq
The forms $\Omega^i_+$ form the basis for the symplectic form $J_\bbb$ and the holomorphic two-form $\omega_\bbb$ on our four-dimensional space \cite{bbw}\footnote{Our conventions differ from those of \cite{bbw}, where the trio of forms defining the hyper-K\"ahler structure is taken to be anti-self-dual while the normalizable (1,1)-forms are self-dual.}. \\

We write the metric on the six-dimensional fibration $X$,  as
\beq
\label{intmetric}
 \mbox{d}s_6^2  = e^{2\Delta} ( e^{-2 u} \mbox{d}s_\bbb^2 +  \Theta \otimes \bar{\Theta}) \, ,
\eeq
%$\Theta^I$ ($I=1,2$) are the well-defined connection one-forms.  
%For the Eguchi-Hanson space there exists a single normalizable real $(1,1)$ form and we can take this to be the curvature of the $\mathbb{T}^2$ bundle. 
%Having constructed a pair of potentials $B^I$, we can choose coordinates on $\mathbb{T}^2$, $\theta^I$, and 
 %We now construct a fibration $\mathbb{T}^2 {\rightarrow} {\tilde X} \stackrel{\pi}{\rightarrow} \bbb$  using 
%the following complex connection
%\beq\label{theta}
%\Theta = \Theta^1 + i \Theta^2 \, .
%\eeq
where $\Theta^I =  \d \, \theta^I + B^I$ ($I=1,2$) are smooth connection one-forms  and we introduced the complex 
one-form $\Theta = \Theta^1 + i \Theta^2$. The connections $\Theta^I$ have
anti-self-dual curvature $F^I \in H^2(\mathcal{B}, \mathbb{Z})$:
\beq\label{asd-bu}
\pi^* F^I = \d \Theta^I \, , \qquad * F^I = -F^I \, .
\eeq
In the complex structure defined by $\omega_\bbb$, these two-forms are of $(1,1)$ type. 
Moreover, they are orthogonal to the self-dual two forms $\Omega^i_{+}$
\beq
F^I \wedge J_\bbb = F^I \wedge \omega_\bbb = 0 \, .
\eeq
%It is useful to define a complex   and a complex curvature $F=F^1+ i F^2$. 

As shown in \cite{GP}, for non vanishing $F$, the  six-dimensional manifold $X$ does not admit a K\"ahler metric. It is not hard to verify  that on $X$ 
%with metric $\ref{intmetric}
%obtained from 
%the fibration $\tilde X$ via scaling ($\d s_6 = e^{2\Delta}  \d s_{\tilde X}$),  
the  fundamental form and complex three-form
\bea
\label{su3x}
J = e^{2\Delta} \, (e^{-2 u } J_\bbb + \frac{i}{2}  \,  \Theta \wedge \bar{\Theta}) \qquad
\qquad \mbox{and} \qquad \qquad  \Omega = e^{3\Delta - 2 u}\,  \,  \omega_\bbb \wedge \Theta
\eea
define an $SU(3)$ structure compatible with the metric (\ref{intmetric}). 
Since the curvature $F$ has no component in  $\Lambda^{0,2} T^*\bbb$, %the curvature of  $\Theta$, 
the fundamental form in \eqref{su3x} obeys
\bea
\label{dj}
\d \, J &=& \d (2 \Delta - u ) \wedge J \nn \\
&-& e^{2 \Delta} [ \d u   \,  \w ( e^{-2u} \, J_{\mathcal{B}} -  \Theta^1 \wedge \Theta^2) +  (\pi^*F^1 \wedge \Theta^2 - \pi^*F^2 \wedge \Theta^1 ) ] \, ,
\eea 
where the second piece corresponds to the primitive part of $\d  J$, and it is not hard 
to check that its wedge product with both  
$J$ and $\Omega$ vanishes. Similarly, on can check that the holomorphic three-form is conformally closed 
\beq
\label{do}
\d (e^{2 u - 3 \Delta} \, \Omega) = 0 \, .
\eeq

As it is clear from the equations above, the six-dimensional  
manifold $X$ is complex but is not symplectic.
%\footnote{Let us recall that, for an $SU(3)$ structure manifold the derivative
%of the $J$ and $\Omega$ can be expanded into torsion classes
%\bea
%&& \d J =\frac{3}{2} {\rm Im}( \bar{W}_1 \wedge \Omega) + W_4 \wedge J + W_3 \, \\
%&& \d \Omega = W_1 \wedge J^2 +W_2 \wedge J + \bar{W}_5 \wedge \Omega \, .  
%\eea
%Then in this case the  non-zero torsion classes are $W_4$ and $W_3$.}. 
Moreover, wedging \eqref{dj} with $J$, we obtain
\beq
\label{balance}
J \wedge \d \, J = \d (2 \Delta - u ) \wedge J^2  \, ,
\eeq
which tells that the metrics considered here  are (conformally) balanced. \\

\subsection{The supersymmetry equations}
\label{eqs}

A convenient way to repackage the supersymmetry conditions in type II supergravity is in terms of pure spinors on the internal manifold.
These are polyforms, obtained as tensor products of the supersymmetry parameters on the internal manifold
\beq
\Psi_\pm =  \frac{8 }{|a|^2} e^{-\phi} \,   e^{-B} \, \eta_+^1 \oplus \eta_\pm^{2 \, \dagger} \, , 
\eeq
where $|a|^2 = ||\eta_+^1||^2 =  ||\eta_+^1||^2$. 
Then the supersymmetry  equations become \cite{gmpt1,gmpt2} 
\begin{subequations}
\bea
\label{int}
\d (e^{3A}\Psi_-)&=& 0 \, , \\
 \label{nonint1}
\d (e^{2A} \Re \Psi_+)&=& 0 \, , \\ 
\label{nonint2}
\d (e^{4A} \Im \Psi_+)&=&  e^{4A} e^{-B} * ( F_1 - F_3 + F_5) \, , 
\eea 
\end{subequations}
where $\phi$  is the dilaton, $B$ is the NS two-form and $F_1$, $F_3$, $F_5$ the RR fluxes on the
internal manifolds. In these equations we have already used the fact that supersymmetry sets the norm of the spinors to be
proportional to the warp factor, $|a|^2= e^A$. So we see that supersymmetry requires that
one of the pure spinors must be closed, while the second one is not integrable due
to the presence of the RR fluxes.\\

For an $SU(3)$ structure manifold the pure spinors $\Psi_{\pm}$ are given by 
\beq
\label{pure}
\Psi_+ = e^{i\theta_+} e^{-\phi} e^{-B} e^{-iJ} \qquad \quad \mbox{and}
\qquad \quad   \qquad \Psi_- = -i \, e^{i\theta_-} e^{-\phi} e^{-B} \, \Omega \, 
\eeq
where $J$ and $\Omega$ are the fundamental form and the holomorphic three-form defining the $SU(3)$ structure.
The choice of one phase is arbitrary\footnote{The $SU(3)$ structure is equivalently defined by a globally defined,
nowhere vanishing chiral spinor on the internal manifold. If we call such a spinor $\eta_+$, then the two supersymmetry
parameters are related to $\eta_+$ by
\beq
\eta_+^1 = e^{A/2} e^{i \theta_a} \eta_+ \qquad \qquad  \eta_+^2 = e^{A/2} e^{i \theta_b} \eta_+
\eeq
and the phases of the pure spinors are given by $\theta_\pm =  \theta_a \mp \theta_b$.}. In this paper we fix them in such a way
to be consistent with the interpolation solution of \cite{bgmpz}
\beq
\theta_+ = w + \frac{\pi}{2} \qquad \qquad  \theta_- =- \frac{\pi}{2} \, ,
\eeq
where $w$ is the interpolating function. \\

Expanding equations  (\ref{int}) and \eqref{nonint1}  into forms of definite degree
yields the following equations for the NS sector
\begin{subequations}
\bea
\label{1f}
&& \sin w \, \d (2 A - \phi) = - \d (\sin w)  \, ,\\
\label{3f}
&& \cos w e^{-2 A} \, \d (e^{2 A} J) + e^{\phi} \, \d (e^{- \phi} \cos w) \, J
= - \sin w \, H \, , \\
\label{5f}
&& \sin w \, J \w \d J   =  \cos w  \, H \w J \, ,\\
\label{4f}
&& \d (e^{3A - \phi } \, \Omega) = 0 \, , \\
\label{6f}
&& \Omega \w H = 0 \, ,
\eea
\end{subequations}
and  expanding  \eqref{nonint2} yields for the RR fluxes
\begin{subequations}
\bea
\label{F1}
* F_1 e^{\phi}  &=&  e^{\phi} \,   \d (e^{-\phi} \cos w ) \wedge J^2/2 \, , \\
\label{F3}
* F_3 e^{\phi} &=& - \sin w  e^{-2 A}\, \d(e^{2 A} J) + \cos w  \, H \, ,\\ 
\label{F5}
* F_5 e^{\phi} &=&    \, e^{-4A+\phi} \d ( e^{4A-\phi} \cos w )\, .
\eea
\end{subequations}

Equation \eqref{4f} comes for the closure of $\Psi_-$, and indicates that the complex structure 
is integrable.  This allows to use the Dolbeault operator in taking the Hodge star, so that 
the RR fluxes can be written as
\begin{subequations}
\bea
\label{F1D}
F_1 &=&   i ( \del - \delbar) (\cos w \, e^{-\phi}) \qquad \Rightarrow \qquad \delbar \left(C + i \cos w \, e^{-\phi} \right) = 0\, ,  \\
%\label{F3D}
%F_3  &=& - i \sin w  \left( e^{-2 A} \, (\partial - \bar{\partial}) (e^{2A-\phi}J) - e^{-2\phi} J \wedge(\partial - \bar{\partial})  e^{\phi}  \right)- e^{-\phi} \cos w * H \, , \\
\label{F5D} 
F_5 &=&   i e^{-4A} ( \del - \delbar) (\cos w \, e^{4A -\phi}) \wedge J^2 /2 \, ,
\eea
\end{subequations}
where $C$ is the RR zero form potential.  So the presence of a holomorphic combination of the
axion and the dilaton is a generic feature of all these backgrounds.  In this paper we shall
be interested in the solutions where the axion is constant. 

The expression for the RR three-form \eqref{F3} can be simplified as well, but the result depends on the value of $w$.  For $w=\pi$,  it reduces to
\beq
e^{\phi}F_3 =  *H \, .
\eeq
When $w \neq \pi$, it is not hard to show that  the two form $e^{(2A - \phi)} J$ is primitive.  This allows to simplify the expression for RR three-form flux 
to 
\beq
\label{F3D}
F_3  = - i \sin w  \left( e^{-2 A} \, (\partial - \bar{\partial}) (e^{2A-\phi}J) - e^{-2\phi} J \wedge(\partial - \bar{\partial})  e^{\phi}  \right)- e^{-\phi} \cos w * H \, . 
\eeq
Note that in our conventions, the $(2,1)$ primitive form is imaginary self-dual. \\

For generic values of the phase $w$  the system of equations \eqref{1f}-\eqref{F5} describes a solution of an ``interpolating" type. 
The special cases of $w= \pi \, (\mbox{mod} \, \pi)$ and $w = \pi/2 \, (\mbox{mod} \, \pi)$ are often referred to as type B and type C
solutions in the literature. Type C solutions are characterised by a vanishing axion and NS three-form, $C=0$ and $H=0$.
On the contrary, type B solutions have an holomorphic dilaton \eqref{F1D} which reduces to the standard complex field  
$\tau = C + ie^{-\phi}$, and we recover the imaginary self-dual three-form $F_3  + \tau H$. Moreover  $H \w J = 0$  and $J$ is conformally closed. \\
 
As always, the supersymmetry equations are only necessary condition for supersymmetric vacua. In order for a background to be a solution, it
must also satisfy the Bianchi identities for the fluxes (the equations of motion are implied by supersymmetry)
\begin{subequations}
\bea
&& \d F_1 = 0 \, , \\
&& \d F_3  - H \wedge  F_1  = 0 \, ,\qquad \qquad  \d H =0 \, ,\\
&& \d F_5 - H \wedge  F_3  = 0 \, .
\eea
\end{subequations}

\section{D5-brane solutions (type C)}
\label{typeC}

In this section we discuss solutions of type C, namely with $w = \pi/2$.
A generic feature of this class of solutions is that the  internal manifold is complex but not K\"ahler. \\

For $\cos w = 0$, the ansatz \eqref{su3x} combined with
\eqref{1f}, \eqref{5f} and \eqref{4f} gives
\beq\label{scal}
2 A = 2 \Delta = \phi =u  \, .
\eeq
In order to simplify notations, here and in the following, we set most of the integration constants to one.
They can be easily reintroduced as constant parameters multiplying the various terms in the metric.

As for the other equations, \eqref{3f}, \eqref{F1} and \eqref{F5} imply
\beq
H = 0 \,  \qquad F_1 = F_5 = 0 \, , 
\eeq
with the only non-vanishing flux given by \eqref{F3}
\beq
\label{3fluxC}
\d \, (e^{2A} J) = - e^{4A} * F_3 \, .
\eeq
Using the integrability of the complex structure \eqref{4f}, it is easily checked that
\beq
\label{3fluxC*}
 F_3  =  i (\partial - \bar{\partial}) (e^{-2A}J) \, .
\eeq
\vspace{0.2cm}

Let us consider type C solutions where the internal manifolds are torus fibrations of
the type discussed in Section \ref{tg}. The internal metric is then
\beq
\label{intmetricC}
 \mbox{d}s_{6\, C}^2  =  e^{- \phi} \mbox{d}s_\bbb^2 + e^{\phi}   \Theta \otimes \bar\Theta \, ,
\eeq
and  the tadpole  condition reads
\beq
\label{tadpoleC}
\d  F_3 =  - 2i \partial \bar{\partial} (e^{-2A} J) =  - 2i \partial \delbar (e^{-2 \phi } J_{\mathcal{B}} + \frac{i}{2} \Theta \wedge \bar{\Theta}) \, .
\eeq
Using the closure of the fundamental form on the hyper-K\"ahler base and the curvature of the torus fibration,
the absence of tadpoles is equivalent to the following equation
\beq
\label{tadpoleC0}
%2 i \partial\bar\partial (e^{- 2 A} J) = 
2 i \partial\bar\partial e^{-2 \phi} \wedge J_\bbb - F \wedge \bar F  \, =\, 0 \, ,
\eeq
which is the only equation left to solve. \\

In solving the supersymmetry constraints we used the fundamental form and 
complex three-form defined in \eqref{su3x}
\bea
\label{su3x2}
J = e^{- \phi } J_\bbb +  e^{\phi }  \frac{i}{2}  \,  \Theta \wedge \bar{\Theta} \, ,
\qquad \qquad  \Omega = e^{-\phi/2}\,  \,  \omega_\bbb \wedge \Theta \, .
\eea
It is easy to see that, by a change of complex structure on the base $\bbb$,  
we obtain a second $SU(3)$ structure 
\bea
\label{su3x3}
J =  - e^{- \phi } J_\bbb +  e^{\phi }  \frac{i}{2}  \,  \Theta \wedge \bar{\Theta} \, 
\qquad \mbox{and} \qquad  \Omega = e^{-\phi/2}\,  \,  \bar \omega_\bbb \wedge \Theta
\eea
which solves all the supersymmetry conditions with the same metric and RR flux.  This means that the solution preserves ${\cal N} =2$ supersymmetry.  

%In the following, we shall refer to the fundamental form \eqref{su3x2} at type C point as $J_C$.
%and $\Omega_C$ ($J_C = J = e^{-2 \phi } J_\bbb + \frac{i}{2} e^{\phi} \,  \Theta \wedge \bar{\Theta})$).   

\subsection{Multi-center  Taub-NUT solutions}

We can find solutions of the tadpole conditions \eqref{tadpoleC0} where the base metric 
\eqref{basemetric} is of Gibbons-Hawking type. These are obtained choosing the function $V$ in \eqref{basemetric} as 
\beq
\label{potbm}
V = \eta + \sum_{q=1}^N \frac{1}{r_q} \, ,
\eeq
where the index $q$ denotes the individual centers and $r_q=|\vec{y}-\vec{y}_q|$ is the distance 
away from a given center, $y_q$.  
For  $\eta=1$, the metric corresponds to  a regular multi-center Taub-NUT . For $\eta=0$, we have 
the ALE spaces. For the special values $q=1$  and $q=2$,  the base is  flat space
or an Eguchi-Hanson space, respectively. \\

As discussed in Section \ref{tg} we need to construct a $\mathbb{T}^2$ fibration with anti-self dual $(1,1)$ curvature.
This can be done using the anti-self-dual two forms $ \Omega^i_-$ in \eqref{2forms}.
Indeed, for a choice of (three-dimensional) harmonic  functions $K^I$ and one-form $\xi^I$ (we shall take $I=1,2$), satisfying the (anti)monopole equation  \cite{bbw}
\beq
\label{anti-monopole}
\partial K^I = +   \epsilon_{ijk} \partial_j \xi^I_k \, ,
%\qquad (\d K^I = -  *_3 \d \xi^I) \, ,
\eeq 
the potentials
\beq
\label{b-pot}
B^I = V^{-\frac{1}{2}}   K^I \, e^0 + \xi^I
\eeq
will yield anti-self-dual field strengths 
\beq
F^I \, = \d \, B^I = - \partial_i (V^{-1} K^I) \Omega^i_- \,  .
\eeq  
Then the torus fibration is defined by 
$\Theta^I= d \theta^I + B^I$. 

On a Taub-NUT space with $N$ centers there are $N-1$ normalizable anti-self dual $(1,1)$ forms 
corresponding to the $N-1$ independent compact two-cycles. Regularity and normalizability require that the positions of the poles of the harmonic functions $K^I$ coincide with the centers 
of the Taub-NUT. The function
 \beq
 K = K^1+ i K^2  =  \sum_{q=1}^N \frac{k_q}{r_q}
 \eeq
 defines a smooth fibration with complex curvature  $F = F_1 + i F_2 \, = - \partial_i (V^{-1} K) \Omega^i_-$.
 %{\bf A choice of  $k_i$ should corresponds to a non compact two-cycle - a disk going to infinity. It should be $k_i=1$. Is that non-normalizable?}. 
 \\
 
The three-form in this case is
\beq
F_3 =  i (\partial - \bar{\partial}) (e^{-2 \phi}) \wedge J_\bbb + \frac{1}{2} (\Theta \wedge \bar{F} + \bar{\Theta} \wedge F ) \, ,
\eeq
and the  tadpole condition (\ref{tadpoleC0})
\beq
\label{tadpoleTN}
2 i \partial\bar\partial e^{-2 \phi} \wedge J_\bbb - F \wedge \bar F =0  
\eeq 
is equivalent to 
\beq
\label{lapl}
\Box_\bbb (e^{-2 \phi}) =  *  ( F \wedge \bar F ) \, .
\eeq
When  the function $\phi$ only depends on the $y^i$ coordinates, \eqref{lapl} reduces to the differential equation 
\beq
\frac{\nabla^2 e^{-2 \phi} }{V} +  \Big | \vec{\nabla} \frac{K}{V} \Big |^2\,  = \, 0 \, ,
\eeq
where derivatives are taken with respect the $y_i$ variables in $\mathbb{R}^3$. The tadpole equation  can be explicitly solved by
\beq\label{solTN}
e^{-2 \phi}  = - \frac{ K \bar K}{2 V} + L
\eeq
with $L(y_i)$ an arbitrary harmonic function on $\mathbb{R}^3$. Since $K/V$ is smooth 
by definition,
the first term on the right hand-side has simple poles at the positions of the Taub-NUT centers. 
All these
singularities can be cancelled and the expression made positive by a suitable choice of coefficients in 
$L = l_0 + \sum_q \frac{l_q}{r_q}$. 
The resulting metric and dilaton are everywhere regular. 

These are  solutions with non trivial fluxes and no branes. By allowing poles in $e^{-2 \phi}$  we could find 
more general solutions with $\d F_3 \neq 0$   corresponding to  D5-branes wrapping the two-torus.

\subsection{The Eguchi-Hanson solution}

Setting  $\eta=0$ in the potential \eqref{potbm}, we obtain the family of  ALE spaces. It is 
interesting to see the explicit expression for the solution in the simplest case of ALE, the Eguchi-Hanson space $T^*P_1(\mathbb{C})$.
It corresponds to  the resolution of  $\mathbb{C}^2/\mathbb{Z}_2$, where the singular point is replaced
by a two-sphere. A similar solution has been recently discussed in \cite{fty} in the heterotic framework. \\

Following \cite{fty}, we use a different set of coordinates with respect to the previous section, which fully exploits the $SO(3)$ invariance of the metric.

The K\"ahler form and holomorphic two-form on the smooth Eguchi-Hanson are given by
\bea
\label{EH}
J_{\mathcal{B}} &=&  \frac{i}{2} \partial\bar \partial K(r^2) = \frac{i}{2} \left[   \frac{1}{r^2} \sqrt{r^4 + a^4}  \del \delbar  r^2  -  \frac{a^4 }{r^4 \sqrt{r^4 + a^4} }\del r^2 \wedge \delbar r^2 \right] \, , \nn \\
\omega_\bbb &=& \d z_1\wedge \d z_2 \, ,
\eea
where $r^2 = |z^1|^2 + |z^2|^2$, the K\"ahler potential is 
\beq 
K(r^2) =\sqrt{a^4+r^4} + a^2 \log \left ( \frac{r^2}{a^2 +\sqrt{a^4+r^4}}\right ) \, ,
\eeq  
and  $a$ is a non-negative constant controlling the size of the two-sphere.  \\

For the Eguchi-Hanson space there exists a single normalizable anti-self dual closed $(1,1)$ form. It corresponds to
the curvature of a line bundle on the two-sphere. The explicit expression has been constructed  in \cite{fty} and it is given by $i \partial\bar\partial \log h$ with
\beq
\frac{\d}{\d r^2} \log h  = \frac{1}{r^2 \sqrt{a^4 +r^4}} \, .
\eeq 
We can take this to be the curvature of the $\mathbb{T}^2$ bundle. If we introduce complex coordinates on the torus, $\theta = \theta^1 + i \theta^2$, then the one-form connection is chosen
\beq
\Theta = \d \theta - i c \partial \log h
\eeq
in such a way that $F = \d \Theta = \bar \partial \Theta$. 
$c$ is a quantized constant, essentially the first Chern number of the torus fibration. Then the three-form flux \eqref{3fluxC*}
reads 
\beq
\label{F3flux}
F_ 3 = \frac{1}{2}  \left[ \, - \, \frac{\d}{\d r^2}(e^{-2 \phi}) \frac{\sqrt{r^4 + a^4}}{r^2} \,  \del \delbar  r^2  \wedge (\partial - \bar \partial) r^2  
+ \Theta \wedge \bar{F}  + \bar{\Theta} \wedge F \right] \, .
\eeq

As before we consider solutions corresponding to pure fluxes. In this case the tadpole condition, $\d F_3 =0$, reduces to 
\beq
\label{fty-tadpole}
2i \partial \delbar (e^{-2 \phi} J_{\mathcal{B}} + \frac{i}{2} \Theta \wedge \bar{\Theta}) = \frac{1}{r^2} 
\frac{\d}{\d r^2} ({\cal J}(r^2) r^4) \d z^1 \wedge \d {\bar z}^1 \wedge \d z^2 \wedge \d {\bar z}^2 \, ,
\eeq
where 
\beq
\label{fty-j}
{\cal J}(r^2) = - \frac{\d}{\d r^2}(e^{- 2 \phi}) \frac{1}{r^2} \sqrt{r^4 + a^4}  + \frac{|c|^2}{r^4 (a^4 + r^4)} \, .
\eeq
The general solution of the tadpole condition is $ {\cal J}(r^2) =\gamma/r^4$, which leads to 
\beq
\label{solC1}
e^{-2 \phi} = \frac{|c|^2}{a^4 \, \sqrt{a^4 + r^4}} + \frac{a^4 \gamma - |c|^2}{a^6}
\, \log [\frac{2 a^4 ( a^2 + \sqrt{a^4 + r^4})}{(a^4 \gamma - |c|^2) r^2} ] + e^{-2 \phi_\infty} \, .
\eeq
This expression is singular at $r=0$ for generic values of the constant $\gamma$. However
for the special value $\gamma = |c|^2/a^4$ the solution reduces to a regular one 
\beq
\label{solC2}
e^{-2 \phi}= \frac{|c|^2}{a^4 \, \sqrt{a^4 + r^4}} +  e^{-2 \phi_\infty}\, .
\eeq

Note that for generic values of $\phi_\infty$ the warp factor approaches a constant for large $r$. In the same limit, the curvature $F$ goes to zero and 
the torus fibration becomes trivial. We obtain a solution which is an asymptotic unwarped ALE space times a two torus. On the other hand, for small $r$,
the non-trivial fibration changes the topology of the internal space. In particular, one of the two torus circles combines with the two-sphere  
in the Eguchi-Hanson space to give a non-trivial three-cycle\footnote{In fact, the Eguchi-Hanson space becomes $\mathbb{R}^2\times S^2$ and one cycle in the two-torus is fibered over $S^2$ with the connection proportional to that of 
the Hopf fibration.}  \cite{GP}. Thus for small $r$, topologically, the space becomes $S^3\times \mathbb{R}^2\times S^1$.  
Since $F^1$ and $F^2$ are proportional, a one-cycle remains non-trivial \cite{GP}.\\

The presence of $F_3$ suggests a brane interpretation. Indeed the solution is similar in spirit 
to the Maldacena-Nunez solution and its generalization \cite{CNP} recently  discussed in \cite{mm}. 
We can even take a near-brane limit by sending  $e^{-2\phi_\infty}$ to zero. For large $r$, the metric becomes a warped product, 
\beq
\d s^2 \, \sim \, r \left ( \d s^2_4 + \d \theta \otimes \d \bar\theta \right ) + \frac{1}{r}\,  \d s^2_\bbb \, ,
\eeq
and the three-form flux reduces to
\beq 
F_3 \, \sim   \, \frac{1}{r^4} \,  \del \delbar  r^2  \wedge (\partial - \bar \partial) r^2 \, .
\eeq
It is easy to check that this expression is the volume form of the three-sphere at infinity in 
the Eguchi-Hanson space.
We thus obtain a warped solution which, asymptotically, represents
the near-horizon geometry of a  D5 brane wrapped on the two-torus.  On the other hand, the two-torus 
disappears in the IR part of the solution. As in a geometric transition, the brane is replaced by a 
flux supported on the non-trivial three-cycle. Indeed, the first term in (\ref{F3flux}) vanishes and 
the three-form reduces to
\beq
F_3 \, \sim \,  \Theta \wedge \bar{F} + \bar{\Theta} \wedge F \, ,
\eeq
which is indeed supported on the IR three-sphere. 

It is important to notice that a non-trivial fibration
is crucial for obtaining a regular solution for small $r$. A D5 brane solution in flat space or on $\mathbb{C}^2/Z_2$ is singular in the IR.  Even the  resolution of the orbifold to a smooth ALE space still leads to a singular solution, as can be seen from equation
(\ref{solC1}) in the limit $c\rightarrow 0$. However, for $c\ne 0$ we obtain a change in topology that allows for a regular
solution.

The large value of the dilaton suggests that we should better use the  S-dual picture representing 
an NS brane. For large $r$  the S-dual solution behaves as 
\bea
\d s_{NS}^2 \, &\sim & \,  \d s^2_4 + \d \theta \otimes \d \bar\theta \, +\, \frac{ \d r^2}{r^2} \,  + \,  \d s^2_{S^3/\mathbb{Z}_2} \, , \nn \\
e^{\phi_{NS}} &\sim &\frac{1}{ r} \, ,
\eea
which represents a linear dilaton background. This is the holographic dual of a little string theory living on a five-brane 
sitting at $\mathbb{C}^2/Z_2$ and compactified on a torus \cite{Aharony:1998ub}.
%{\bf holographic interpretation of $a$ and $c$?}

A similar analysis can be performed for the general solution based on Taub-NUT spaces. 
%A general discussion of the topology of the resulting six-dimensional manifold is given in \cite{GP}.

\subsection{Breaking supersymmetry to ${\cal N}=1$}

There is a simple generalization of the solutions obtained above by adding a $(2,0)$ term to the torus curvature
\beq
\d \Theta = F + G \, ,
\eeq
where $G\in H^{(2,0)}(\bbb,\mathbb{Z})$. The total curvature $F+G$ is still orthogonal to the fundamental and holomorphic two-forms on $\bbb$
\beq
(F+G) \wedge J_\bbb = (F+G) \wedge \omega_\bbb = 0 \, ,
\eeq 
and this is enough to solve equations (\ref{5f}) and (\ref{4f}) 
and all the other supersymmetry constraints.
Since a reverse in the complex structure of $\bbb$ will transform a $(2,0)$ form into a $(0,2)$ one, this
solution will preserve only one of the two supersymmetries.

The general form of the new piece is $G = g \, \omega_\bbb$, where $g$ is an arbitrary holomorphic function 
on $\bbb$, which can be non-trivial since the space is non compact. 
For simplicity we analyse the solution with $\bbb$ given by  the
Eguchi-Hanson space and $g$ constant. In this case it is easy to see what happens to the tadpole condition,
which becomes
\bea
2i \partial \delbar (e^{-2 \phi} J_{\mathcal{B}} + \frac{i}{2} \Theta \wedge \bar{\Theta}) &=& 2i \partial \delbar (e^{-2 \phi}) \wedge J_{\mathcal{B}} - F\wedge \bar F + G \wedge \bar G  \nn \\
&=& 2i \partial \delbar (e^{-2 \phi}) \wedge J_{\mathcal{B}} +  ( |F|^2 + |G|^2) \mbox{Vol}_4 \, ,
\eea
since $\bar \partial \Theta = F$ and $\partial \Theta = G$.  Using (\ref{EH}), we see that
the tadpole equation has the same form as in \eqref{fty-tadpole} with  an
additional constant contribution to the function $\mathcal{J}(r^2)$
%\beq
%\label{fty-tadpole2}
%2i \partial \delbar (e^{-2 \phi} J_{\mathcal{B}} + \frac{i}{2} \Theta \wedge \bar{\Theta}) 
%= \frac{1}{r^2} 
%\frac{\d}{\d r^2} ({\cal J}(r^2) r^4) \d z^1 \wedge \d {\bar z}^1 \wedge \d z^2 \wedge \d {\bar z}^2 \, ,
%\eeq
%with 
\beq
\label{fty-j2}
{\cal J}(r^2) = - \frac{\d}{\d r^2}(e^{- 2 \phi}) \frac{1}{r^2} \sqrt{r^4 + a^4}  + \frac{|c|^2}{r^4 (a^4 + r^4)}  -\frac{|g|^2}{2} \, .
\eeq
The regular solution of the tadpole condition $ {\cal J}(r^2) =\gamma/r^4$ is now
\beq
\label{solC20}
e^{-2 \phi}= \sqrt{a^4 + r^4} \left ( - \frac{ |g|^2}{2} + \frac{c^2}{a^4 \, ( a^4 + r^4 )}\right ) +  \delta \, .
\eeq
This expression changes sign and it is not well behaved. The presence of a $(2,0)$ piece in the curvature unpleasantly changes the asymptotic behavior of the solution.  \\

It would be of some interest to explore further the possibility of having regular $\mathcal{N}=1$ solutions.

\section{T-duality and D3-brane solutions (type B)}
\label{typeB}

Examples of type B solutions with $w=\pi$ can be obtained from the type C backgrounds discussed above via T-duality. In general, type B solutions can have a holomorphic dilaton.
A particular class is obtained for constant dilaton and vanishing axion. The metric is conformally Calabi-Yau and the
(complex) three-form flux is imaginary self-dual. \\

%For simplicity we will consider the Eguchi-Hanson solution, even if most of the results hold for the general Taub-NUT case.
Since the original type C-solution  has two isometries, corresponding to shifts in the angles $\theta^I$ on the torus, 
we can perform two T-dualities along those directions. The effect of T-duality is to untwist the torus fibration and to produce a $B$ field
with one leg along the torus. The type B dilaton is constant while the original type C dilaton is related
to the type B warp factor. Indeed, a standard application of the T-duality rules gives 
\begin{subequations}
\label{typeBsol}
\bea
&& \d s^2_6 = e^{ - 2 A} (\d s^2_\bbb + \d \theta \otimes \d \bar{\theta}) \, ,\\
&& e^\phi = g_s  \, ,\\
&& A = \phi_C/2 \, ,\\
&& B = B^1 \wedge \d \theta^1+  B^2 \wedge   \d \theta^2  \, ,
\eea
\end{subequations}
where $B^I$ are the potentials defined in \eqref{b-pot} and the function $\phi_C$ denotes 
the dilaton of the type C solution. This was determined as the solution
of the tadpole equation of type C, \eqref{solTN} for the general Taub-NUT solutions and \eqref{solC2} for the Eguchi-Hanson case. Notice that in type B the warp factor satisfies the equation
\beq\label{tadpoleB}
\Box_\bbb (e^{- 4 A}) =   *   (F \wedge \bar F) \, .
\eeq

%Since the former is supposed to have a conformally Ricci flat metric, the $T^2$ fibration there is trivial, while all components  of $H$ have one leg along the fiber and are given by primitive two-forms on the base. 
%Indeed the contraction of $H$ with a holomorphic vector field on the two-torus is
%\beq
%\iota_{\del_\theta} H = a_I F^I =  H^{2,-}(\mathcal{B})
%\eeq
%with $\theta = \theta^1 + i\theta^2$. After the T-duality, $H$ vanishes and instead 
%we have a non-trivially fibered two torus with connection whose curvature is
%\beq
%F \in H^{2,-}(\mathcal{B}) \, .
%\eeq
%The RR sector  just follows the lead.

The result of T-duality can be summarized in the standard picture of the correspondence space \cite{bem}. 
On one side, we have a non trivial fibration
\beq
\mathbb{T}^2 {\rightarrow} X_{\rm C} \stackrel{\pi}{\rightarrow}\bbb
\eeq
with a vanishing $H$ (which is parametrized by a trivial two-torus). On the other,
the manifold is  a direct product 
\beq
X_{\rm B} = \bbb \times \mathbb{T}^2 \, ,
\eeq
and the geometrisation of the non trivial $H$ yields another  nontrivial two-torus fibration 
over the base $\bbb$. The correspondence space is a fiberwise product  
$X_{\rm B} \times_\bbb  {X_{\rm C} }$, 
and is a double $\mathbb{T}^2$ fibration over $\bbb$ with one of the fibrations being trivial 
and the the other having a connection with curvature $F$. The roles of these two tori 
are reversed in going from type B solution to type C. \\

\begin{figure}[ht!]
\begin{equation} \label{correspondenceb}
\xymatrix @=5pc @ur 
{ \hspace{-1cm}{ \mbox{(type C)} \,\,  X_{\rm C}} \ar[d]_{ \tilde \pi} & 
{ X_{\rm B} \times_\bbb  {X_{\rm C} } } \ar[d]_{p} \ar[l]^{\tilde p} \\ \bbb & {\qquad \quad X_{\rm B} \,\, \mbox{(type B)}} \ar[l]^{  \pi}} \nonumber
\end{equation}
\caption{ {\it The solutions B and C are related via standard T-duality and can be obtained  respectively via projections $\tilde p$ and $p$ form the correspondence space.  $\tilde \pi$ and $\pi$ are the projections of dual torus bundles to base $\bbb$. }}\label{xy}
\end{figure}
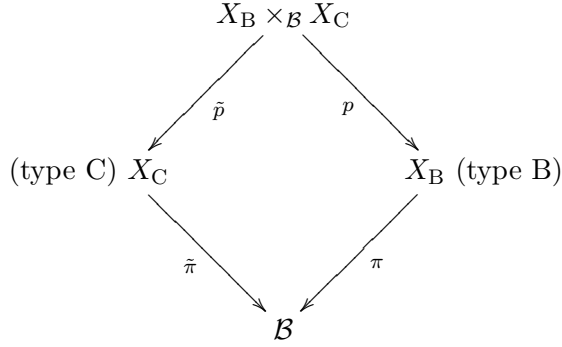

The RR fluxes for the T-dual solution can be obtained from those  of type C by
\beq
\label{RRTd}
\left. e^{-B} F \right|_{\rm type B} = (\iota_{\theta^1} + \d \theta^1) \cdot  (\iota_{\theta^2} + \d \theta^2) \left. (e^{-B} F) \right|_{\rm type C} \, ,
\eeq
where $F=F_1 + F_3 +F_5$ is the sum of the RR field strength on the internal manifold. In this case we obtain
\bea
\label{RRB}
F_1 &=&  0 \, , \nn \\
g_s F_3 &=&  - F^1 \wedge \d \theta^2 +  F^2 \w \d \theta^1  \, , \nn \\
g_s F_5 &=&  i (\partial - \bar \partial)(e^{-4 A}) \wedge J_\bbb \wedge \d \theta^1 \wedge \d \theta^2 \, ,
\eea  
where $g_s$ is the constant value of the dilaton.

From \eqref{RRTd} one sees that the Bianchi identities for the RR fields of the type B solution are automatically satisfied
if the type C ones are\footnote{Bianchi identities  $\d (e^{-B} F) =0$  are automatically satisfied when a shift in $\theta^i$ is an isometry (${\cal L}_{\theta^i}  (e^{-B} F) = 0$ ) since $\d$ and the operators $\iota_{\theta^i} + \d \theta^i$ anticommute. For the T-duality conventions we refer to \cite{gmpw}.}. \\

The solution has an interpretation in terms of D3-branes. Let us discuss for simplicity the case of the Eguchi-Hanson solution.
In the near brane limit, $e^{-2\phi_\infty}\rightarrow 0$, the metric asymptotes at large $r$
\beq
\d s^2 \, \sim \, r \, \d s^2_4 + \frac{1}{r}  \left (\d \theta \otimes \d \bar\theta  +  \d s^2_\bbb \right ) \, ,
\eeq
corresponding to D3-branes smeared on the two-torus. The metric remains a trivial product of the ALE space with a torus also at smaller
values of $r$. On the other hand, we have an imaginary self dual flux for the three-forms. Notice that the $F_5$ flux 
vanishes for small values of $r$. Solutions of this type have been
discussed in \cite{torinesi}, in the case where the two-torus is replaced by $\mathbb{R}^2$, and have been
interpreted as a combination of physical and fractional D3 branes 
probing a resolved $\mathbb{C}^2/Z_2$ singularity. 
%The three-form flux accounts for the distribution of fractional branes on $\mathbb{R}^2$ and the 
The vanishing
of physical D3 brane charge in the IR is typical for these kinds of solution. 
The solutions of \cite{torinesi} are singular in the IR. 
We see that, in the case where we compactify two directions, it is possible to find 
a regular solution. 
%Notice that the background has an {\bf enhanced} ${\cal N}=2$ supersymmetry.
\\

Finally let us discuss the T-duality transformation on pure spinors. 
The type C solution has a trivial $B$-field, and thus the generalized tangent bundle is simply given by $TX_{\rm C} \oplus T^*X_{\rm C}$. The pure spinors are 
globally defined and using \eqref{scal} can be written as 
\bea
\label{puretypeC}
\left. \Psi_+ \right|_{\mbox{type C}} &=&  - e^{-\phi_C}  \exp \left[-i( e^{- \phi_C }J_{\mathcal{B}} +  \frac{i}{2} \, e^{\phi_C}\, 
\Theta \wedge \bar{\Theta})\right] \, ,\nn \\
% \qquad \mbox{and} \qquad 
\left. \Psi_- \right|_{\mbox{type C}}  &=& - \, e^{- 3 \phi_C/2}\,  \omega_\bbb \wedge \Theta \,.
\eea
Note that since the axion is vanishing in this solution, only four-manifolds with globally defined $\omega_\bbb$, namely the hyper-K\"ahler ones, can be 
admissible bases. Since the pure spinors are also polyforms,  T-duality act on them as in \eqref{RRTd}. Then
T-duality along the two-torus  maps \eqref{puretypeC} into a pair of type B pure spinors: 
\bea
\label{puretypeB}
\left. \Psi_+ \right|_{\mbox{type B}} &=&  - i e^{-B - i J} \, ,\nn \\
\left. \Psi_- \right|_{\mbox{type B}} &=& - e^{-B} \wedge \Omega \, ,
\eea
with the fundamental form and the holomorphic three-form given by
\bea
&& J = e^{- 2 A} (J_{\mathcal{B}}  +   \d \theta^1 \wedge \d \theta^2 ) \, , \nn \\
&& \Omega = i e^{- 3 A} \omega_\bbb \wedge (\d \theta^1 + i \, \d \theta^2)  \, ,
\eea
since $A= \phi_C/2$.

% 
%\begin{figure} 
%\begin{equation} \label{correspondenceb}
%\xymatrix @=5pc @ur 
%{ \hspace{-1cm}{ \mbox{(type C)} \,\,  X_{\rm C}} \ar[d]_{ \tilde \pi}  \hspace{0.2 cm} \ar@{-->}[dr]_{
% \pi/2 \leq w <\pi} &
%{ X_{\rm B} \times_\bbb  {X_{\rm C} } } \ar[d]_{p} \ar[l]^{\tilde p} \\ \bbb & {\qquad \quad X_{\rm B} \,\, \mbox{(type B)}} \ar[l]^{  \pi}} \nonumber
%\end{equation}
%\caption{\textit{The solutions B and C are related via standard T-duality and can be obtained  respectively via projections $\tilde p$ and $p$ form the correspondence space.  $\tilde \pi$ and $\pi$ are the projections of dual torus bundles to base $\bbb$. For non-compact $\bbb$ type C and B solutions can be also related via an interpolation with  $ \pi/2 \leq w <\pi$ . For vanishing axion, the family is smooth, except for type B point $w= \pi$. }}\label{xy}
%\end{figure}

%.\footnote{The condition of integrability of complex structure \eqref{4f} can be can be rewritten as
%\beq
%\label{omega-f}
%\d (e^{i \alpha} \, \Omega)  +  \left(i (Q - \frac{1}{4} e^{\phi} \d \tau)  - \frac{3}{4} \d(4A - \phi) \right)  \w (e^{i \alpha} \, \Omega)= 0 \, ,
%\eeq
%where  $Q =  \frac{1}{4} e^{\phi} \d C = \frac{1}{4} e^{\phi} F_1$.}

\section{An interpolating solution: D5-D3 solutions}
\label{interpo}

While finding solutions of B and C type is relatively simple,  solutions corresponding to generic values of the parameter $w$ are
harder to find. A general discussion of the features of interpolating solutions was given in \cite{FG}. A notable regular example  is provided by the
solution \cite{bgmpz} describing the baryonic branch of the Klebanov-Strassler  solution, which interpolates between Klebanov-Strassler 
(type B) and Maladacena-Nu\~{n}ez (type C). Another interesting (although singular)  example is given by the gravity dual of non commutative D5 branes, where the interpolation parameter is given by the value of a $B$ field \cite{Hashimoto:1999ut,Maldacena:1999mh}.
Here we present an interpolating solution  which is regular and combines features of both these examples.
The solution can be obtained in two different ways
\begin{itemize}
\item starting from point C, we can apply the prescription of \cite{mm}, which, through  a chain of dualities, generates  
a family of interpolating solution for {\it any}  type C solution with only three-form flux
\item starting from point B, we can add a constant $B$ field  and, using the isometries of the background, T-dualize along $\mathbb{T}^2$.
\end{itemize}
Before entering into the details,
we discuss some general features of interpolating solutions with vanishing axion.

As already discussed in Section \ref{tg}, the most general solution with $ \pi/2 < w < \pi$ 
has a conformally balanced metric ($J^2$ is conformally closed), the three forms are given by
%Indeed it can be shown that
%\beq\label{prim}
%J \w \d J =  \cot w \, \d w \, J \w J  = - \d (2A- \phi) \w J \w J \, ,
%\eeq 
\begin{subequations}\label{HF3}
\bea
\label{HD}
H &=&  - \d (\cot w \, J) \, ,  \\
\label{F33D}
F_3 &=&  -\frac{i}{\sin w}\, e^{-2A}  ( \del - \delbar) (e^{2A -\phi} J) +  i \, {\sin w} \, e^{-\phi} \,(\del - \delbar)(2A) \w J \, ,
\eea
\end{subequations}
and the Bianchi identity for $H$ trivially follows from \eqref{HD}.

If we also choose to have constant axion, the supersymmetry equations in Section \ref{eqs} 
simplify and, most notably, from 
\eqref{F1} and \eqref{1f} we have 
\begin{subequations}\label{const}
\bea
\d (e^{-\phi} \, \cos w) &=& 0 \,  , \\
\d (e^{2A - \phi} \, \sin w) &=& 0 \, ,
\eea
\end{subequations}
respectively. These equations can be integrated to 
\beq
\label{mm-par}
\cos w = - e^{\phi - \phi_{\infty}} \tanh \beta \, , \qquad  \qquad \sin w = \frac{e^{-\phi_{\infty}}}{\cosh \beta} e^{-(2A - \phi)} \, ,
\eeq
and, as consequence, the warp factor reads
\beq
\label{warp}
e^{-4 A}  = 1 + \cosh^2 \beta (e^{-2(\phi - \phi_{\infty})} -1)  \,.
\eeq 
To simplify comparisons with the existing literature, we used for the integration constants the parametrisation suggested in \cite{mm}.  $\phi_\infty$ is  the asymptotic value of the dilaton and $\beta$ is the boost parameter entering the chain of dualities in \cite{mm}. Notice that we normalized the asymptotic value of the warp factor to one; a generic value can be easily reintroduced.  

Finally, a generic feature of interpolating solutions, which follows straightforwardly from the supersymmetry equations, is that the RR three-form
is always proportional to the Hodge star of the NS one
\beq
F_3 = - \frac{e^{-\phi}}{\cos w} * H \, .
\eeq
This relation generalises the imaginary self-duality condition of type B solutions.

\subsection{The chain of  ``external" dualities}

It was shown in \cite{mm} that, 
starting from a type C solution with only three-form
flux, it is possible to generated a family of backgrounds of interpolating type satisfying all supersymmetry constraints.  All these solutions have zero axion as the original type C case. 
More precisely, given any type C solution with internal metric $\d s^2_{6 \, C}$, fundamental form $J_C$ and dilaton $\phi_C$, we obtain an interpolating solution of the form
\begin{subequations} \label{mm-metric}
\bea
&& \d s^2 = e^{2A} \d s_4^2 + e^{-2 A + \phi} \d s^2_{6 \, C}  \, \\ 
&& \phi = \phi_C \, ,
\eea
\end{subequations}
and fluxes
\begin{subequations}\label{fluxmm}
\bea
H &=&   \sinh \beta \, \d (e^{\phi_C}  J_C) \, \\
F_3  &=&  e^{\phi_\infty} \, \cosh \beta \, e^{- 2 \phi_C}  * \d ( \, e^{\phi_C}  J_C) \,  , \\
F_5 &=&  - \frac{i}{2} \sinh 2 \beta \, e^{  \phi_{\infty}}  (\del - \delbar) (\phi_C) \wedge J_C^2   \, .
\eea
\end{subequations}

In \cite{mm} this family was derived from the type C solution via  a chain of dualities involving a lift to M theory and an eleven-dimensional boost.
It is straightforward to verify that the above expressions satisfy all supersymmetry constraints.

We can turn to the Bianchi identities now.  The closure $H$ is satisfied trivially. A remarkable feature of this solution is that the Bianchi identities for $F_3$ and $F_5$ reduce to those for the type C solution\footnote{In 
deriving \eqref{tad2} it is useful to use the primitivity of the form $e^{2A-\phi} J$ (and in particular that $\del (e^{2A-\phi} J) \wedge \delbar (e^{2A-\phi} J) = -  e^{2A-\phi} J \wedge \del \delbar (e^{2A-\phi} J)$. Note that since $e^{-2 A_C}J_C$ is not primitive, the solution of \eqref{tad2} requires $ \del \delbar (e^{-2 A_C} J_C) = 0$. Remember also that, in type C, 
$2 A_C = \phi_C$.}
\begin{subequations}\label{tadpole}
\bea
\d F_3 = 0  &\Rightarrow&  \del \delbar (e^{-2 A_C} J_C) = 0 \, , \\ \label{tad1}
\d F_5 - H \wedge F_3 = 0  &\Rightarrow&  \del \delbar (e^{-2 A_C} J_C) \wedge J_C= 0 \, . \label{tad2}
\eea
\end{subequations}
Notice from \eqref{fluxmm} that the  new $F_3$ is proportional to the type C one 
given by \eqref{3fluxC}. \\

In \cite{mm}, this construction was used to recover the baryonic 
branch solution of \cite{bgmpz} from the pure type C solution of \cite{CNP}. However this construction is more general and can be applied to our torsional backgrounds as well. 

The interpolating solutions for our Taub-NUT  type C backgrounds are
 obtained from the above equations with the metric and fundamental forms defined
in Section \ref{typeC}. In particular
\beq
J_C \, = \, e^{- \phi_C } J_\bbb +  e^{\phi_C }  \frac{i}{2}  \,  \Theta \wedge \bar{\Theta} \, .
\eeq
%The interpolating metric is still a non trivial torus fibration over the Taub-NUT base, there are D3 and D5 fluxes 
%and a running  dilaton. 

Finally, the pure spinors are given by (in terms of the type C dilaton $\phi=\phi_C$):
\bea
\label{pureint}
\left. \Psi_+ \right|_{\mbox{int}} &=&   i e^{-\phi_{\infty}}
\frac{ i e^{-2A} - \sinh \beta}{\cosh \beta} \,   \exp \left[-(\sinh \beta + i e^{-2A}) ( J_{\mathcal{B}} +  \frac{i}{2} \, e^{2\phi}\, \Theta \wedge \bar{\Theta})\right] \, ,\nn \\
\left. \Psi_- \right|_{\mbox{int}}  &=& i \, e^{- 3A}\,  \omega_\bbb \wedge \Theta \,.
\eea
The relation between the dilaton and the warp factor is given by \eqref{warp}.

\subsection{The ``internal"  dualities}

The same solution can be obtained from  the type B background of Section \ref{typeB} by a particular $O(2,2)$ transformation consisting of adding a closed $B$ field and  T-dualizing along $\mathbb{T}^2$.\\

Let us start from a type B solution with arbitrary parameters, including the volume of $\mathbb{T}^2$, which 
was previously set to one for simplicity. The solution is
\bea
&& \d s^2_6 = e^{ - 2 A_B} (\d s^2_\bbb +  R^2 \d \theta \otimes \d \bar{\theta}) \, ,\nn \\
&& e^\phi = g_s  \, ,\nn \\
&& B = B^1 \wedge \d \theta^1 +  B^2 \wedge \d \theta^2 \, ,  
\eea
where $A_B$ is a solution of the tadpole equation (\ref{tadpoleB})
\beq
\Box_\bbb (e^{- 4 A_B}) =  \frac{1}{R^2}  *   (F \wedge \bar F) \, ,
\eeq
where the rescaling of the $\theta^I$ has been taken into account.
%which has the form given in \eqref{solTN} for the general Taub-NUT solutions and \eqref{solC2} for the Eguchi-Hanson %case.  
%Note that, with $R\ne 1$, there is a corresponding rescaling in the tadpole equation.

%\bea
%&& J = e^{- 2A} (J_{\mathcal{B}}  +   e^{-2\phi_{\infty}} \frac{i}{2 \cosh^2 \beta}  \, \d \theta \wedge \d \bar{\theta} ) \, , \nn \\
%&& \Omega = e^{- \frac{3}{2} A}  e^{-\phi_{\infty}} \frac{1}{ \cosh \beta} \,\omega_\bbb \wedge \d \theta  \, ,
%\eea
%and 
%\beq
%e^{-\phi} = e^{-\phi_{\infty}}\frac{1}{\cosh \beta} \, .
%\eeq
%The NS three-form flux is given by:
%\beq\label{new-h}
%H = \frac{1}{2} \Re ( F \wedge \d {\bar \theta}) \, .
% \eeq

We may perform now a constant $B$-transform, an operation that while shifting the $B$-field will not change the flux $H$
\beq
\label{new-bfield}
B \longrightarrow B +  b\, \d \theta_1\wedge \d \theta_2 +  d\,  J_\bbb \,. 
\eeq

The crucial term in this redefinition is the constant two-form along the torus, which does not commute with double T-duality. The base-dependent part in the $B$-transform proportional to $J_\bbb$ has been included in order to be able to collect the result of T-duality onto the form \eqref{pureint}.\\

A standard application of T-duality along $\theta^1$ and $\theta^2$ now gives the interpolating solution. 
In the previous subsection, various arbitrary parameters, including the radius of the torus and the asymptotic value of the warp factor, have been set to one. In order to recover the same normalizations  we need to choose $g_s = R$, $d=-b/R^2$ and the asymptotic value of $e^{-4 A_B}$ equal to one. Obviously, we could relax these conditions
and obtain interpolating solutions with arbitrary parameters. Note that the warp factor for the interpolating solution is the same as in type B ($A=A_B$) while the dilaton  is given by the standard rules of T-duality
\beq
e^{-2\phi} = \frac{1}{g_s^2} \left ( b^2 + R^4 e^{-4 A}\right ) = \frac{1}{R^2} \left ( b^2 + R^4 e^{-4 A}\right ) \, ,
\eeq
from which, by comparison with (\ref{warp}), we read the value of the parameters $\beta$ and $\phi_\infty$
\beq
e^{-2 \phi_\infty} = \frac{b^2 +R^4}{R^2} \,\, \qquad\qquad \,\, \sinh \beta =  - \frac{b}{R^2} \, .
\eeq
Since $e^{-2\phi}$ and $e^{-4 A}$ are linearly related, they both satisfy the standard tadpole equation.
We can choose $\phi=\phi_C$ of the form \eqref{solTN} for the general Taub-NUT solutions and \eqref{solC2} for the Eguchi-Hanson case. $e^{-4 A}$ has a similar form with different coefficients; in particular, 
for us, its asymptotic value is normalized to one. \\

%This is easier to verify on the pure spinors, using that the T-duality is given by $T = (\d \theta^1 + \imath_{\partial_{\theta^1}}) (\d \theta^2 + \imath_{\partial_{\theta^2}})  = b +{\hat  \beta} + a $, where the $B$ and $\beta$ transforms and the $SL(2)$-rotation are given respectively by $b = \d \theta^1 \wedge \d \theta^2$, ${\hat \beta}= \imath_{\partial_{\theta^1}} \wedge \imath_{\partial_{\theta^2}}$ and $a = \d \theta^1 \wedge \imath_{\partial_{\theta^2}} - \d \theta^2 \wedge \imath_{\partial_{\theta^1}}$. 

%It looks like the same thing might be obtained without any tuning of the constants, but  an  $SL(2)$ transformation of the complexified volume of $\mathbb{T}^2$ actiong as $\tau=  B_{\theta_1 \theta_2} + i \det(g_{\theta_1 \theta_2})   \rightarrow \frac{\tau}{1+ \gamma \tau}$  with 
%$$
%\gamma e^{-\phi} = \tan w \, , \qquad 1 + \gamma e^{-\phi} = e^{\phi_{\infty}} \coth \beta\,  e^{-\phi} e^{i w}
%$$
%but things do not quite work out!

There exists another $O(2,2)$ transformation that relates  the type B background with the interpolation \eqref{pureint}.
We could start with the type B solution with general parameters and perform a Lunin-Maldacena transformation (LM) \cite{LM} 
\beq
\left ( \begin{array}{cc} 1 & 0 \\ \gamma & 1 \end{array} \right ) \in SL(2, \mathbb{Z})
\eeq
acting on the complexified K\"ahler parameter of the two-torus.  It is easy to check that we obtain again the interpolating solution up to a constant shift of the final $B$ field. The continuous parameter $\gamma$ plays the role of $b$ in the
T-duality transformation. The two parameters $\phi_\infty$ and $\beta$ are related by a  redefinition to the torus radius
$R$ in the type B solution  and the  value of $\gamma$.

Alternatively, one may start with a  D3 branes transverse to $\bbb \times \mathbb{T}^2$ without three form fluxes, and combine a non-constant $B$ transform with twisting of the two-torus in such a way that the original integrable complex structure does not change. The transformed even pure spinor $\Psi_+$ will generate the correct three-form fluxes. In \cite{amp} such a transformation was used to relate compact type B and C backgrounds. In our case the novelty is that the  the phase of the pure spinor does not change in a discrete fashion.

\subsection{Properties of the interpolating solution}
\label{abeh}

In this section we will discuss some features of the interpolating solutions described above.
For simplicity, we will focus on the simple case of the Eguchi-Hanson base.

The ten-dimensional metric becomes 
\beq
\label{mm-metric1}
\d s^2 = e^{2A} \d s_4^2 + m \, e^{- 2 A } ( \d s_\bbb^2 +   e^{2 \phi_C} \, \Theta \otimes \bar{\Theta}) \, ,
\eeq
with dilaton ($\phi=\phi_C$) and warp factor %\eqref{warp}
\bea
\label{dilwarp}
e^{-2 \phi_C} &=& \frac{|c|^2}{a^4 \, \sqrt{a^4 + r^4}} + e^{-2\phi_{\infty}} \, , \nn \\
e^{-4 A}  &=& 1 + \cosh^2 \beta (e^{-2(\phi_C - \phi_{\infty})} -1)  \, .
\eea

The fluxes are
\bea
\label{solut}
F_3 &=&  i m \, e^{\phi_\infty} \, \cosh \beta \, (\del - \delbar) (e^{- 2 \phi_C}  J_\bbb + \frac{i}{2} \Theta \wedge \bar{\Theta})  \, ,\nn \\
H  &=&  \frac{i m}{2}  \sinh \beta \,  \d (e^{2 \phi_C} \Theta \wedge \bar{\Theta} ) \, ,\nn \\
F_5 &=&   \frac{1}{2} m^2 \sinh 2 \beta \, e^{  \phi_{\infty}}  (\del - \delbar) (\phi_C) \wedge J_\bbb \wedge \Theta\wedge \bar\Theta \, .
\eea
We introduced an arbitrary constant $m$ which will be useful very soon.\\

For generic $\beta$, the warp factor $A$ goes to a constant for large $r$, as seen from (\ref{dilwarp}). 
As noticed in \cite{mm}, we can also obtain a near-brane solution by sending $\beta\rightarrow\infty$ \footnote{This observation was used
to obtain the the gravity dual of the baryonic branch of the Klebanov-Strassler theory \cite{bgmpz} (which is a near-brane solution) starting from  the generalization of the Maldacena-Nunez solution given in \cite{CNP}.}.
%The general construction of \cite{mm} gives a family of backgrounds with asympotically constant warp factor and the baryonic branch is then obtained by 
%sending $\beta\rightarrow\infty$. 
In order to have a smooth limit, we need to rescale the space-time coordinates and the parameter $m$,
\beq
\label{resc}
ds^2_4 \, \rightarrow \frac{|c| \, e^{\phi_\infty} }{a^3} \cosh \beta \,\, ds^2_4 \,\, , \qquad\qquad m \, \rightarrow \frac{a^3}{|c| \, e^{\phi_\infty} \, \cosh \beta }  \, .
\eeq
Factors of $c,a$ and $e^{\phi_\infty}$ are introduced for future convenience.
In this way, we obtain 
\bea
\label{mm-metric2}
\d s^2 &=& e^{2A} \d s_4^2 +  \, e^{- 2 A } ( \d s_\bbb^2 +   e^{2 \phi_C} \, \Theta \otimes \bar{\Theta}) \, ,\nn\\
e^{-2 \phi_C} &=& \frac{|c|^2}{a^4 \, \sqrt{a^4 + r^4}} + e^{-2\phi_{\infty}} \, , \nn \\
e^{-4 A}  &=& 
%\frac{a^6}{c^2 e^{2\phi_\infty}} \left (e^{-2(\phi_C - \phi_{\infty})} -1\right )  = 
\frac{a^2} {\sqrt{a^4 +r^4}} \, , \nn \\
F_3 &=&  i \, \frac{a^3}{|c|}\,  (\del - \delbar) (e^{- 2 \phi_C}  J_\bbb + \frac{i}{2} \Theta \wedge \bar{\Theta})  \, ,\nn \\
H  &=&  i \,  \frac{a^3}{2 |c|} \, e^{-\phi_\infty} \d (e^{2 \phi_C} \Theta \wedge \bar{\Theta} ) \, ,\nn \\
F_5 &=&   \frac{a^6}{|c|^2} \, e^{ - \phi_{\infty}} (\del - \delbar) \phi_C \wedge J_\bbb \wedge \Theta\wedge \bar\Theta \, . 
\eea
Differently from Section \ref{typeC}, where we obtained a near-brane solution by sending the asymptotic value of the dilaton to zero, here $\phi_\infty$
has an arbitrary value and the near-brane limit is obtained by sending $\beta$ to infinity. 

%This solution has many similarities both with non-commutative D5 branes solution and with the gravity dual of the baryonic branch of the Klebanov-Strassler theory \cite{BGMPZ}. 
Notice first that the solution  is still of interpolating type and is  a mixture of those discussed in Sections \ref{typeC} and \ref{typeB}.  
For large $r$ the metric behaves as
\beq
\d s^2 \, \sim \, r \, \d s^2_4 + \frac{1}{r}  \left (   \d s^2_\bbb  + e^{2 \phi_\infty} \d \theta \otimes \d \bar\theta  \right ) \, ,
\eeq
and there is  a non-vanishing $B$ field along the two-torus. The asymptotic solution corresponds to a D5 brane with
a non commutative parameter on the world-volume \cite{Hashimoto:1999ut,Maldacena:1999mh}.  The $F_3$ flux is supported on the three-sphere at infinity of the ALE space. The $B$ field induces a non trivial
D3 charge and the metric looks like that of D3 branes smeared on the two-torus. The torus shrinks for large $r$ but the metric is regular.  
%As discussed in the literature, at least for rational B, an $SL(2,\mathbb{Z})$ transformation
For small $r$, as in the type C solution,
the internal manifold becomes topologically $\mathbb{R}^2\times S^1\times S^3$, the D3 brane charge vanishes 
and the $F_3$ flux is supported by the IR three-sphere. Notice that the dilaton is running but approaches a constant both for large and small $r$. All this is reminiscent of the gravity dual of the baryonic branch of the Klebanov-Strassler theory \cite{bgmpz}.
 
In the case of the baryonic branch, the resulting solution is de facto an {\it interpolation} between type C and type B solutions. Indeed, with suitable tuning of parameters, 
the solution connects the Klebanov-Strassler point (type B) with the Maldacena-Nunez solution (type C). 
Interestingly, a similar phenomenon happens for our solutions; by varying parameters, we can interpolate between the near-brane type B solution of Section \ref{typeB} and the near-brane type C solution of Section \ref{typeC}. We can conveniently use $\phi_\infty$ and the value $\phi_0$ of the dilaton at $r=0$ as parameters. We will keep $\phi_0$ and $a$ fixed and we will vary $\phi_\infty$.
We have a consistent solution for all $\phi_\infty \ge \phi_0$.  $c$ is then determined as 
\beq
|c| = a^3 \sqrt{e^{-2 \phi_0} -e^{-2 \phi_\infty}} \, .
\eeq
For $\phi_\infty\rightarrow\phi_0$, we have that $c\rightarrow 0$ and the metric becomes a direct product for all
values of $r$\footnote{Without factors of $c$ in the rescaling  (\ref{resc}), we would have obtained a type B case by sending $c\rightarrow 0$ without varying $\phi_\infty$. However equations (\ref{mm-metric2}) imply a constant warp factor $A$ and so an uninteresting Calabi-Yau  solution consisting of an ALE space times a two-torus and no flux.} 
\bea
\d s^2 &=& e^{2 A} ds^2_4 + m e^{-2 A} \left ( \d s^2_\bbb + e^{2\phi_0} \d \theta \otimes \d \bar\theta \right )\, , \nn \\
e^{-4 A} &=& \frac{a^2}{\sqrt{a^4+r^4}} \, .
\eea
This is the same as the B solution discussed in Section \ref{typeB}. A short calculation shows that, in the limit $c\rightarrow 0$, also the fluxes are consistent with a type B solution with constant dilaton.
On the other hand, for $\phi_\infty\rightarrow\infty$ , $2 A\sim \phi_C$, the $H$ and $F_5$ fluxes vanish and we
recover the type C solution of Section \ref{typeC} with an effective value of the parameter $c= a^3 e^{-\phi_0}$. 
All metrics in the interpolation have the same  asymptotic behavior. At the endpoint
$\phi_\infty \rightarrow\infty$ the dilaton blows up at infinity
\beq
e^{-2 \phi} \sim \frac{1}{\sqrt{a^4+r^4}} \, .
\eeq
%changing the asymptotic behavior. 
Again this is similar to what happens for the baryonic branch solution \cite{bgmpz}. \\
%We thus obtain an interpolation by varying from $c=0$ (the type B of Section (\ref{typeB}) to $c=\infty$ (the type C
%solution of Section \ref{typeC}). 
%Notice that the parameter $\phi_\infty$ is kept fixed during the interpolation. It corresponds to the type B constant value of the dilaton. It should not be confused
%with the parameter $d$ appearing in Section \ref{typeC}; indeed in our interpolating family the C point has a blowing up dilaton corresponding to a parameter $d=0$
%in the notation of Section \ref{typeC}.

\section{Discussion and conclusions}

In this paper we constructed new interpolating solution of type IIB based on non compact fibration of a two-torus
over a Taub-NUT space. 

In the ${\cal N}=2$ case the resulting family of solution is regular. We discussed in great details the case of the
Eguchi-Hanson base where the solution and its topology can be studied quite explicitly. In the general solution, the asymptotic value of the warp factor is constant, corresponding to an asymptotically 
unwarped background, but we also considered a near-brane limit. 
The backgrounds depend on many parameters, including the asymptotic values of the dilaton and warp factor, and the details of the fibration.
Some of these parameters will be quantized. This is certainly the case for the Chern class of the fibration, $c$.
In addition, since we have non-trivial three cycles in the geometry, the periods of the NS and RR three-forms lead 
to other quantization conditions. 

Since our solutions are non compact, the question of a possible gauge dual arises naturally. 
Let us consider first the simple case of the type C solution, corresponding to a five-brane wrapped on a two-torus
near an ALE space.  The dual gauge theory should come from the compactification on the torus of the
six-dimensional gauge theory living on a D5-brane near an ALE space. 
When the tours is replaced by flat space ($c=0$)  and $a=0$, the solution is the holographic dual of 
a little string theory associated to a $\mathbb{C}^2/Z_2$ singularity. At low energy it reduces 
to a six-dimensional gauge theory coupled to tensor multiplets with eight supercharges,  two gauge groups
and two bi-fundamental hyper-multiplets. In the case of a more general ALE space with $N$ centers, the theory is based on an affine $A_N$ quiver coupled to tensor multiplets, with $N$ gauge groups and bi-fundamental hyper-multiplets connecting neighbouring groups \cite{Intriligator:1997dh}. 
In this picture, the parameter $a$ corresponds to a resolution of the singularity and it is usually associated with a baryonic VEV in the gauge theory. 
The parameter $c$ is less straightforward to interpret and is related to the compactification of the theory on the torus.
It should correspond to a twist of the theory by some global $U(1)$ in such a way to preserve ${\cal N}=2$ supersymmetry. 

In the interpolating solution, a non-commutative parameter is also turned on. 

What is interesting and unusual is that, in spite of being $\mathcal{N}=2,$ the dual of such gauge theory is  regular and, in the IR,
is topologically similar to the known IR behaviour of ${\cal N}=1$ confining theories with a non-trivial three-sphere that supports
the RR flux. In contrast  with the MN and KS case, there is an extra circle, surviving  the twist, where we could wrap strings.    
%At least in the regime where a near brane interpetation is valid, the solutions should correspond to D5-branes wrapping the torus-fiber. 
%{\bf Can we say anything about the gauge theory? the difference in asymptotics between $\N=2$ and $\N=1$? the meaning of the EH (or general Taub-NUT) parameters on the gauge theory side (holographic interpretation of $a$ and $c$?)? ...} 
It would be interesting to understand the role of such strings in the IR dynamics and to understand better the structure and moduli space of the $\mathcal{N}=2$ dual theory. We will leave these
issues for future work.\\

\section* {Acknowledgements}

We would like to thank I.~Bena, S.~Giusto, L.~S.~Tseng  and N.~Warner for useful discussions.  We would also like to thank  KITP, UC Santa Barbara (RM) and  GGI Center of Physics
in Florence  (MP and AZ) for hospitality during the course of this work. This work is supported in part by RTN contracts  MRTN-CT-2004-005104 and  MRTN-CT-2004-512194 and by ANR grants BLAN05-0079-01 (DA and MP) and BLAN06-3-137168 (RM). A.~Z.~ is supported in part by INFN and MIUR under contract 2007-5ATT78-002.

\end{document}